\documentclass[twocolumn,pre,floats,showpacs,floatfix]{revtex4}
\usepackage[final]{epsfig}
\usepackage{amsmath, amsthm, amssymb}
\usepackage{color}

\newcommand{\beq}{\begin{equation}}
\newcommand{\eeq}{\end{equation}}
\newcommand{\beas}{\begin{eqnarray*}}
\newcommand{\eeas}{\end{eqnarray*}}
\newcommand{\bea}{\begin{eqnarray}}
\newcommand{\eea}{\end{eqnarray}}

\newcommand{\bdar}{\leftrightarrow}

\long\def\symbolfootnote[#1]#2{\begingroup\def\thefootnote{\fnsymbol{footnote}}\footnote[#1]{#2}\endgroup}

\begin{document} 
\title{Reciprocity of Networks with Degree Correlations and Arbitrary Degree Sequences}

\author{Gorka Zamora--L\'opez$^1$} 
\author{Vinko Zlati\'{c}$^{2}$} \email{vzlatic@irb.hr}
\author{Changsong Zhou$^1$}
\author{Hrvoje \v{S}tefan\v{c}i\'{c}$^2$}
\author{J\"urgen Kurths$^1$}

\affiliation{$^1$Institute of Physics, University of Potsdam PF 601553, 14415 Potsdam, Germany}
\affiliation{$^2$Theoretical Physics Division, Rudjer Bo\v{s}kovi\'{c} Institute, 
   	P.O.Box 180, HR-10002 Zagreb, Croatia}
	
\date{\today} 

\begin{abstract}
Although most of the real networks contain a mixture of directed and bidirectional (reciprocal) connections, the reciprocity $r$ has received little attention as a subject of theoretical understanding. We study the expected reciprocity of networks with an arbitrary input and output degree sequences and given 2-node degree correlations by means of statistical ensemble approach. We demonstrate that degree correlations are crucial to understand the reciprocity in real networks and a hierarchy of correlation contributions to $r$ is revealed. Numerical experiments using novel network randomization methods show very good agreement to our analytical estimations. 
\end{abstract}
\pacs{89.75.Fb, 89.75.Hc, 02.10.Ox, 02.50.Cw} 
\maketitle

\section{INTRODUCTION}
Most of real networks combine both unidirectional and bidirectional (reciprocal) connections. This directed nature is often obviated, e.g. networks are symmetrized for algorithmic convenience and network models largely ignore the directionality for analytical simplicity. However, the directionality is known to be relevant, e.g., robustness against environmental changes of metabolic networks seems to arise from evolutionary pressure on the directions and weights of the metabolic fluxes~\cite{Fischer_Fluxanalysis_2005}.
The formation of functional communities and hierarchies in the cerebral cortex is mediated by the presence of reciprocal and unidirectional connections~\cite{Zhou_HierarchFunct_2006}.
The dynamical stability in complex networks, e.g. ecological systems~\cite{Gardner, May} and synchronization of coupled oscillators, is commonly assessed by the eigenvalue space of the Jacobian or Laplacian matrices~\cite{Pecora_MasterStab_1998}. When networks are directed these matrices will have complex eigenvalues, what influences both the stability and the dynamical organization far from the equilibrium state.

The network reciprocity $r$ is classically defined as $r = \frac{L^{\bdar}}{L}$~\cite{WassermanBook} where $L^{\bdar}$ is the number of directed links $s \to t$ that also have a reciprocal (bidirectional) counterpart $s \gets t$, and $L$ is the total number of directed links. In networks without self-loops, reciprocal links form the cycles of lowest order and are therefore, important as a natural measure of feedback in the network. Recently, $r$ of the Wikipedia networks \footnote[1]{Wikipedia networks are constructed out of the published documents under the Wikipedia.org website. Only internal links within the Wikipedia articles are considered.} (for different languages) was found to be very stable over a wide range of network sizes~\cite{Zlatic_Wiki_2006} which signals its relevance for the structure or functionality of the networks.
In~\cite{Serrano_WWW_2005} it was shown that reciprocal connections carry most of the topological information of the WWW. The formation of the giant-component in directed networks is facilitated by reciprocal connections~\cite{Boguna_Percolation_2005}. Despite the extensive modeling efforts during the recent years to reproduce realistic features of networks, models have largely ignored reciprocity. Only in~\cite{Garlas_Grandcanonical_2006} a general class of random networks with prescribed $r$ has been presented.

When analyzing real networks it is important to test whether measured values are significant or not. Typically, real network measures are compared to the properties of complete random networks of the same size $N$ and number of links $L$. However the degree distribution of most real networks largely differs from the aforementioned random networks and more reasonable comparison is desired. Derivation of analytical expressions for expected measures under conditions of arbitrary degree sequence is difficult. Further assumptions are usually introduced, e.g. scale-free~\cite{Schwartz} or exponential degree distributions~\cite{Snijders_2004}. 
As real networks do never exactly belong to a model class, if even close, for real applications a numerical approach is the only solution in most situations. Ensembles of maximally random networks can be generated with the same input and output degree sequences and the ensemble average properties can be calculated. Unfortunately, generation of such ensembles is computationally very demanding for large networks. In this paper, we present analytical expressions for the expected reciprocity of directed networks that can be evaluated using only information measurable from the specific real network under study, and thus, overcome the problems discussed above.

\begin{figure} 
	\centering
	\vspace{-0.2cm}
 	\includegraphics[width=0.47 \textwidth,clip=]{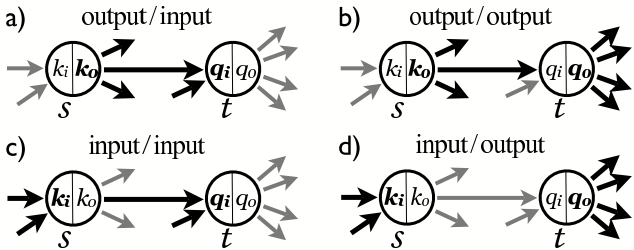}
 	\caption{2-node 2-degree correlations ($2n2d$) of \emph{neighboring nodes} in directed networks. Links corresponding to correlated degrees are colored black.
	\label{fig:knn}}
\end{figure}

In the context of social networks, several redefinitions of $r$ have been introduced that account for biases of the experimental conditions~\cite{WassermanBook, Mandel}. Recently, a similar redefinition based on correlations of adjacency matrix and its transpose, was presented~\cite{Garlas_Reciprocity_2004}. It evaluates the reciprocity with respect to the density of connections $\bar{a}  = L / (N(N-1))$. This results from the fact that for a maximally random network with $N$ nodes and $L$ links, $r$ equals $\bar{a}$. However, the specific degree sequence is expected to affect the number of reciprocal links, i.e., a node with both large input degree $k_i$ and large output degree $k_o$ has a higher tendency to form reciprocal links. This is expressed by the 1-node degree correlations between $k_i$ and $k_o$ of individual nodes. Similarly, as $r$ involves the pairwise connectivity of nodes, the correlation between the degrees of neighboring nodes, 2-node degree correlations should also be relevant. Imagine a link $s \to t$ connecting a source node $s$ with input and output degrees $(k_i,k_o)$ to a target node $t$ with degrees $(q_i,q_o)$. 
The 2-node degree correlations exist when any of the $(k_i,k_o)$ degrees of node $s$ is correlated with any of the $(q_i,q_o)$ degrees of node $t$.
When all four values are correlated, then both 1-node ($1n$) correlations and the 2-node 4-degree ($2n4d$) correlations are present. The class of 2-node 2-degree ($2n2d$) correlations is depicted in Fig.~\ref{fig:knn}. 

In this paper we study the expected reciprocity $\langle r \rangle$ of networks with \emph{prescribed degree sequences and arbitrary $2$-node degree correlations.} We consider complex networks as members of the statistical ensemble with given node degree sequence and degree correlations, and we calculate the expected reciprocity of such ensembles in the thermodynamical limit. We find that degree correlations explain almost completely the observed $r$ of some real networks. In other examples, larger discrepancies indicate the presence of additional internal structure.

\section{GENERAL RESULTS}
In order to analytically estimate $\langle r \rangle$ under different correlation structures, we characterize the real directed networks by: the number of nodes $N$, the number of links $L$,
the number of nodes $N(k_i,k_o) = N(\boldsymbol{k})$ having in-degree $k_i$ and out-degree $k_o$, and the number of directed links $L(\boldsymbol{k} \rightarrow \boldsymbol{q})$ pointing from nodes with degrees $(k_i,k_o)$ to nodes with degrees $(q_i,q_o)$. All these properties are easy to measure in a real directed network and contain all the relevant information about the degree correlations. 
We use frequencies of these properties as their probabilities and calculate the expected number of reciprocal links $\left<L^{\bdar}\right>$. Remind that by definition, $\left< r \right>$ is related to $\left<L^{\bdar}\right>$ by $\left< r \right> = \left<L^{\bdar}\right> /L$.

Under the class of 1-node and 2-node degree correlations here assumed, a network is considered as maximally random when \emph{any of the nodes with degrees $\boldsymbol{k}$ is equally likely connected to any of the nodes with degrees $\boldsymbol{q}$.} If a network contains $L(\boldsymbol{k} \to \boldsymbol{q})$ such links, the probability that any of them connects randomly chosen nodes of degree $\boldsymbol{k}$ and $\boldsymbol{q}$ respectively is, \vspace{-0.2cm}
	\begin{equation} \label{eq:ProbLink}
	p(\boldsymbol{k} \to \boldsymbol{q}) = \frac{L(\boldsymbol{k} \to \boldsymbol{q})}{N(\boldsymbol{k})N(\boldsymbol{q})}. 
	\end{equation}
\noindent in the thermodynamical limit~
\footnote{For thermodynamical limit it is understood the limit of large networks ($N \to \infty$) with sparse connectivity (density $\bar{a} \to 0$). In such a limit the probability of random introduction of multiple links can be discarded.}.
The denominator $N(\boldsymbol{k}) N(\boldsymbol{q})$ is the number of all possible connections between nodes with degrees $\boldsymbol{k}$ and nodes with degrees $\boldsymbol{q}$.

Again, if the network has $L(\boldsymbol{k} \to \boldsymbol{q})$ links of the type $\boldsymbol{k} \to \boldsymbol{q}$, the expected number of reciprocal $\boldsymbol{k} \bdar \boldsymbol{q}$ links is then $\left< L( \boldsymbol{k} \bdar \boldsymbol{q}) \right> = L(\boldsymbol{k} \to \boldsymbol{q}) \, p(\boldsymbol{k}\leftarrow \boldsymbol{q})$.
The overall expected reciprocity $r_{1n2n}$ of the whole network is obtained by summing $\left< L( \boldsymbol{k} \bdar \boldsymbol{q}) \right>$ over all $\boldsymbol{k}$, $\boldsymbol{q}$ degree combinations:
\vspace{-0.2cm}
	\begin{equation}\label{eq:Genr}
	r_{1n2n} =  \frac{1}{L} \sum_{\boldsymbol{k,q}}
\frac{L(\boldsymbol{k} \rightarrow \boldsymbol{q}) \, L(\boldsymbol{k} \leftarrow \boldsymbol{q})} {N(\boldsymbol{k}) \, N(\boldsymbol{q})}.
	\end{equation}

\noindent Note that in general $L(\boldsymbol{k} \rightarrow \boldsymbol{q})\neq L(\boldsymbol{k} \leftarrow \boldsymbol{q})$. Taking frequencies of nodes $P(\boldsymbol{k}) = N(\boldsymbol{k})/N$ and frequencies of links
$\mathcal{P}( \boldsymbol{k} \rightarrow \boldsymbol{q}) = L(\boldsymbol{k} \rightarrow \boldsymbol{q})/L$ as probabilities in the thermodynamical limit, Eq.~(\ref{eq:Genr}) reads,
\vspace{-0.2cm}
	\begin{equation}\label{eq:Genrprob}
	r_{1n2n} = \frac{L}{N^2} \sum_{\boldsymbol{k,q}}
\frac{\mathcal{P}(\boldsymbol{k} \rightarrow \boldsymbol{q}) \, \mathcal{P}(\boldsymbol{k} \leftarrow \boldsymbol{q})} 	{P(\boldsymbol{k}) \, P(\boldsymbol{q})}.
	\end{equation}

\noindent The contribution of the correlation structure is accounted by the sum, $\mathcal{P}(\boldsymbol{k} \to \boldsymbol{q})$ accounts for both 1-node and 2-node correlations, and $P(\boldsymbol{k})$ only for the 1-node correlations. When all four degrees are independent $\mathcal{P}(\boldsymbol{k} \to \boldsymbol{q}) = P(k_i) \, k_o \, P(k_o) \, q_i \, P(q_i)\, P(q_o) / \left<k\right>^2$ where $\left<k\right> = L/N$ is the average degree and $P(\boldsymbol{k}) = P(k_i) \, P(k_o)$. Then reciprocity reduces to the density of links $L/N^2$, i.e., the expected reciprocity of uncorrelated random networks.

\section{SPECIAL CORRELATION CLASSES} \label{special}
Equations~(\ref{eq:Genr}) and~(\ref{eq:Genrprob}) are general formulas that account for all 1-node and 2-node degree correlations. These equations can be reduced to consider only desired special classes of correlations and thus explore the contribution of individual correlation types to $r$. In this section we present detailed derivations for all $8$ possible combinations of 1-node and 2-node correlations and in Table~\ref{tab:analytical} the main results are summarized. Note that  in Table~\ref{tab:analytical} only $r_c$, the contribution of the correlation structure is shown. The expected reciprocities are obtained by multiplying with the density of links: $\left<r\right> = \frac{L}{N^2} \, r_c$ . Along this section, the usual product rule of joint probabilities in terms of conditional probabilities will be used,
\vspace{-0.2cm}
	\begin{eqnarray}\label{ProductRule}
	P(X_1,X_2,...,X_n)&=&P(X_1)\times P(X_2|X_1)\times...\nonumber\\
	&&...\times P(X_n|X_1,X_2,...,X_{n-1}). \nonumber
	\end{eqnarray}

\noindent With this rule in mind, the joint probabilities associated with the link statistics $\mathcal{P}(\boldsymbol{k} \to \boldsymbol{q}) \equiv p(k_i,k_o,q_i,q_o)$ can be expressed, for example, as:
\vspace{-0.2cm}
	\begin{equation}\label{approxOutIn}
	\mathcal{P}(\mathbf{k} \rightarrow \mathbf{q}) = \mathcal{P}(k_o \rightarrow q_i)P(k_i\mid k_o \rightarrow q_i)P(q_o 	\mid k_o \rightarrow q_i, k_i) \nonumber
	\end{equation}

\subsection{2-node out-/in-degree correlations}
Suppose that a network has significant 1-node correlations and 2-node correlations only between the out-degree $k_o$ of source nodes and the in-degree $q_i$ of the target nodes, see Fig.~\ref{fig:knn}(a), while other possible 2-node correlations are negligible. In this case, the probabilities can be approximated by:
\vspace{-0.2cm}
	\begin{eqnarray}
	\mathcal{P}(\boldsymbol{k} \rightarrow \boldsymbol{q}) \approx \mathcal{P}(k_o \rightarrow q_i) \, P(k_i|k_o)\, P(q_o|q_i) \nonumber \\
	\mathcal{P}(\boldsymbol{k} \leftarrow \boldsymbol{q}) \approx \mathcal{P}(k_o \leftarrow q_i) \, P(k_o|k_i)\, P(q_i|q_o) \nonumber
	\nonumber
	\end{eqnarray}

\noindent Using conditional probabilities for the degrees of individual nodes, e.g. $P(k_i|k_o) = P(\boldsymbol{k}) / P(k_o)$, Eq.~(\ref{eq:Genrprob}) reduces to:
\vspace{-0.2cm}
	\begin{equation}\label{eq:1n2noutin}
	r_{1n2n:o/i} = \frac{L}{N^2}\sum_{\boldsymbol{k,q}}\frac{\mathcal{P}(k_o \to q_i) \mathcal{P}(k_i \gets q_o) P(\boldsymbol{k}) P(\boldsymbol{q})} {P(k_i)P(k_o)P(q_i)P(q_o)}.
	\end{equation}

\noindent Additionally, if the 1-node correlations are negligible, the degrees of individual nodes become independent: $P(\boldsymbol{k} ) = P(k_i,k_o) = P(k_i) \, P(k_o)$. Then Eq.~(\ref{eq:1n2noutin}) becomes:
\vspace{-0.2cm}
	\begin{equation} \label{eq:2noutin}
	r_{2n:o/i} = \frac{L}{N^2} \sum_{\boldsymbol{k,q}} \mathcal{P}(k_o \to q_i) \mathcal{P}(k_i \gets q_o) = \frac{L}{N^2},
	\end{equation}
\noindent that equals the density of connections $\bar{a}$ in the thermodynamical limit. This means that the 2-node out-/in-degree correlations do not contribute to reciprocity in the absence of 1-node correlations. This is not a general case since other classes of 2-node correlations largely contribute to $r$.

\subsection{1-node degree correlations}
Starting from Eq.~(\ref{eq:1n2noutin}), we can alternatively remove the remaining 2-node in-/out-correlations and obtain a general expression for the expected reciprocity $r_{1n}$ due to the 1-node correlations alone. In this case, the number of all possible output connections from source nodes with degree $k_o$ is $k_o \, N(k_o)$ and the number of all possible input connections to target nodes with in-degree $q_i$ is $q_i \, N(q_i)$. Hence, the probability that one link connects a node with out-degree $k_o$ to a node with in-degree $q_i$ is $\mathcal{P}(k_o \rightarrow q_i) = \frac{k_oN(k_o)}{L} \, \frac{q_iN(q_i)}{L} = \frac{k_oP(k_o)q_iP(q_i)}{\left<k\right>^2}$. Equation~(\ref{eq:1n2noutin}) reduces to:
\vspace{-0.2cm}
	\begin{equation}\label{eq:1n}
	r_{1n} = \frac{L}{N^2}\frac{\left<k_ik_o\right>^2}{\left<k\right>^4},
	\end{equation}

\noindent Here $r$, a two node property, is determined by the single node characteristics arising from the specific in- and out-degree sequences. We remind that, in the literature, it is common to randomize networks by methods that conserve the degree sequences in order to obtain expected values accounting for the real degree distribution. Equation~(\ref{eq:1n}) is of relevance for significance testing because it is the theoretical estimation of the expected reciprocity in such a typical case.

\begin{table}
\centering
\begin{tabular}{c c @{\quad} l}
\hline\hline
1-node & 2-node & \hspace{0.5cm}$r_c$, contribution of the correlations\\
\hline
no & no & $1$\\
yes  & no & $\frac{\left<k_ik_o\right>^2}{\left<k\right>^4}$\\
\hline \hline
no & out--in & $1$\\
yes & out--in &  $\sum_{\mathbf{k},\mathbf{q}}\frac{\mathcal{P}(k_o \to q_i)\mathcal{P}(k_i \gets q_o)P(k_i,k_o)P(q_i,q_o)}{P(k_i)P(k_o)P(q_i)P(q_o)}$\\
\hline 
y/n & out--out & $\sum_{k_o,q_o}\frac{\mathcal{P}(k_o \to q_o)\mathcal{P}(k_o \gets q_o)}{P(k_o)P(q_o)}$ \\
\hline
y/n & in--in & $\sum_{k_i,q_i}\frac{\mathcal{P}(k_i \to q_i)\mathcal{P}(k_i \gets q_i)}{P(k_i)P(q_i)}$ \\
\hline
no & in--out &  $\frac{(\left<k_iq_o\right>_\mathcal{P})^2}{\left<k\right>^4}$
\symbolfootnote[1]{The averaging in the formulae is performed over in degrees
of source nodes and the out-degree of the target nodes.}\strut\\
yes & in--out &  $\sum_{\mathbf{k},\mathbf{q}}k_ik_oq_iq_o\cdot$\\
&& $\cdot\frac{\mathcal{P}(k_i \to q_o)\mathcal{P}(k_o \gets q_i)P(k_i,k_o)P(q_i,q_o)}{\bar{k}_{i,k_o}\bar{k}_{o,k_i}\bar{q}_{i,q_o}\bar{q}_{o,q_i}}$\symbolfootnote[2]{$\bar{k}_{o,k_i}=\sum_{k_o'}k_o'P(k_o',k_i)$
and similar for all other averages of this type.}\\
\hline \hline
\end{tabular}

\caption{Reduced formulae for the expected reciprocity of different 
combinations
of 1-node and 2-node correlations. The expected reciprocity is $\left<r\right> = \frac{L}{N^2} \, r_c$. Note that the result for the 2-node out--out and in-in correlations are independent of the 1-node correlations.}
\label{tab:analytical}
\end{table}

\subsection{2-node out-/out-degree correlations}
Following a similar approximation we can calculate the expected reciprocity $r_{2n:o/o}$ due to the 2-node correlations, Fig.~\ref{fig:knn}(b), between the output degree $k_o$ of the source node and the output degree $q_o$ of the target node. However, in this case a few steps need to be carefully considered. We rewrite the link probabilities as:
\vspace{-0.2cm}
	\begin{equation}
	\mathcal{P}(\mathbf{k} \rightarrow \mathbf{q})=\mathcal{P}(k_o \rightarrow q_o)P(k_i|k_o \rightarrow q_o)P(q_i|k_i,k_o \rightarrow q_o).
	\nonumber
	\end{equation}

\noindent The term $P(q_i|k_i,k_o \rightarrow q_o)$ introduces dependence of in-degree of the target node $q_i$ on the in-degree of the source node $k_i$. We are assuming that such correlations are negligible and therefore this term can be rewritten as: $P(q_i|k_o \rightarrow q_o)$. The second term $P(k_i|k_o \rightarrow q_o)$ can be written as $P(k_i|k_o)$, because the in-degree of the target node $q_i$ and the in-degree of the source node $k_i$ are, again, not correlated. To proceed, it is necessary to use the fact that we are calculating expectations on \emph{graphs}, and that the final expression should has a form which can be calculated using only the assumed knowledge of the network structure, i.e. the frequencies of nodes $P(\boldsymbol{k}) = N(\boldsymbol{k})/N$ and the frequencies of links $\mathcal{P}(\boldsymbol{k} \to \boldsymbol{q}) = L(\boldsymbol{k} \to \boldsymbol{q})/L$. Following this line of reasoning the third ,already approximated term, should be carefully rewritten. What is the probability that a node will have in-degree $q_i$ given that (i) it has an input link and (ii) it has an out-degree $q_o$? Without the information (i) this probability is simply $P(q_i|q_o)$, but we have information that such a link does exist. The probability that, following a link between nodes with degrees $k_o$ and $q_o$, it will run into a node with in-degree $q_i$ is just the number of links $L(\rightarrow \mathbf{q})$ which enter all nodes with degrees $\mathbf{q}$ divided by the number of all links $L(\rightarrow q_o)$ that enter the nodes with out degree $q_o$. These numbers of links can be expressed as $q_iN(q_i,q_o)$ and $\sum_{q'_i}q'_iN(q'_i,q_o)$ respectively. Thus,
\vspace{-0.2cm}
	\begin{equation}
	P(q_i|k_o \rightarrow q_o) = \frac{q_iN(q_i,q_o)}{\sum_{q'_i}q'_iN(q'_i,q_o)} = \frac{q_iP(q_i,q_o)}{\sum_{q'_i}q'_iP(q'_i,q_o)}
	\nonumber
	\end{equation}

\noindent Substituting all these approximations in Eq.~(\ref{eq:Genrprob}),
\vspace{-0.1cm}
	\begin{equation}
	r_{2n:o/o} = \frac{L}{N^2}\sum_{\substack{\boldsymbol{k,q}}}\frac{\mathcal{P}(k_o \rightarrow q_o)\mathcal{P}(k_o \leftarrow q_o)k_iP(\mathbf{k})q_iP(\mathbf{q})} {\sum_{\substack{k'_i,q'_i}}k'_iP(k'_i,k_o)q'_iP(q'_i,q_o)P(k_o)P(q_o)}
	\nonumber
	\end{equation}

\noindent Finally, the summation terms over in-degrees in the numerator and the denominator cancel out and we obtain a final expression for the expected reciprocity under 2-node out-/out-degree correlations:
\vspace{-0.1cm}
	\begin{equation} \label{eq:2noutout}
	r_{2n:o/o} = \frac{L}{N^2} \sum_{k_o,q_o} \frac{\mathcal{P}(k_o \to q_o) \, \mathcal{P}(k_o \gets q_o)} {P(k_o) \, P(q_o)}
	\end{equation}

\noindent Interestingly, this expression is independent of the 1-node correlations even if we did not explicitly assume it. The expression for 2-node in-/in-degree correlations conserved, Fig.~\ref{fig:knn}(c), is the same only with $k_o$ and $q_o$ replaced by $k_i$ and $q_i$, see Table~\ref{tab:analytical}.

\subsection{2-node in-/out-degree correlations}
We now describe the case in which significant in-out correlations, Fig.~\ref{fig:knn}(d), are present in the network. Such correlations are supposed to influence reciprocity considerably because the probability that a given link of the type $\boldsymbol{k} \rightarrow \boldsymbol{q}$ has a reciprocal counterpart $\boldsymbol{k} \leftarrow \boldsymbol{q}$ is directly proportional to the in-degree $k_i$ of the source node and the out-degree $q_o$ of the target node. Following the previous line of reasoning, the link probability can be expressed in this case as:
\begin{equation}\label{approxInOut}
\mathcal{P}(\mathbf{k} \rightarrow \mathbf{q}) = \mathcal{P}(k_i \rightarrow q_o)P(k_o\mid k_i \rightarrow q_o)P(q_i \mid k_i \rightarrow q_o, k_o),
\nonumber
\end{equation}

\noindent and the conditional terms as:
\vspace{-0.2cm}
	\begin{eqnarray}\label{CondInOut}
	P(k_o\mid k_i \rightarrow q_o)&=&\frac{k_o N(k_o,k_i)}{\sum_{k'_o} k'_oN(k'_o,k_i)} =\frac{k_o P(k_o,k_i)}{\sum_{k'_o} k'_oP(k'_o,k_i)}\nonumber\\
	&=&\frac{k_o P(k_o,k_i)}{\bar{k}_{o,k_i}}, \nonumber \\
	P(q_i\mid k_i \rightarrow q_o)&=&\frac{q_i N(q_o,q_i)}{\sum_{q'_i} q'_iN(q_o,q'_i)} = \frac{q_i P(q_o,q_i)}{\sum_{q'_i} q'_iP(q_o,q'_i)}\nonumber\\
	&=&\frac{q_i P(q_o,q_i)}{\bar{q}_{i,q_o}}. \nonumber
	\end{eqnarray}

\noindent Note that the expression $\bar{k}_{o,k_i}=\sum_{k'_o} k'_oP(k'_o,k_i)$ is \emph{not} the average out-degree of nodes with in-degree $k_i$, because $P(k'_o,k_i)$ is the joint probability and not the conditional probability. The average out-degree of nodes with in-degree $k_i$ is $\left<k_o\right>_{k_i}=\frac{\bar{k}_{o,k_i}}{P(k_i)}$.
Using the above relationships we obtain from Eq.~(\ref{eq:Genrprob}) an expression for the expected reciprocity $r_{1n2n:i/o} = r_{i/o}$ due to 1-node and 2-node in-/out-degree correlations:
\vspace{-0.1cm}
	\begin{eqnarray}\label{InOut1}
	r_{i/o} =  \frac{L}{N^2} \sum_{\substack{\mathbf{k},\mathbf{q}}}\frac{k_ik_oq_iq_o\mathcal{P}(k_i \rightarrow q_o)\mathcal{P}(k_o \leftarrow q_i)P(\mathbf{k})P(\mathbf{q})}{\bar{k}_{i,k_o}\bar{k}_{o,k_i}\bar{q}_{i,q_o}\bar{q}_{o,q_i}}.
	\end{eqnarray} 

\noindent Additionally, if the 1-node correlations are removed, the expected reciprocity $r_{2n:i/o}$ due to the 2-node in-/out-degree correlations alone is:
\vspace{-0.1cm}
	\begin{eqnarray}\label{InOutWithout1N}
	r_{2n:i/o}&=& \frac{L}{N^2} \sum_{\substack{\mathbf{k},\mathbf{q}}}\frac{k_ik_oq_iq_o\mathcal{P}(k_i \rightarrow q_o)\mathcal{P}(k_o \leftarrow q_i)}{\left<k_o\right>_{k_i}\left<k_i\right>_{k_o}\left<q_o\right>_{q_i}\left<q_i\right>_{q_o}}\nonumber\\
	&=&\frac{L^2}{N^2}\frac{\left<k_iq_o\right>^2}{\left<k_i\right>^4}.
	\end{eqnarray} 
\noindent Note that without 1-node correlations $\left<k_o\right>_{k_i}=\sum_{k'_o}k'_oP(k'_o,k_i)/P(k_i)=\left<k_o\right>$.

\section{APPLICATION TO REAL NETWORKS}
The class of degree correlations chosen in this paper is not only very interesting from the theoretical point of view, in this section we pay attention to its practical relevance. First, our results are applied to several real networks and the impact of degree correlations on reciprocity is discussed. Finally, in order to proof the validity of our equations as expectation values, the theoretical estimations are compared to the empirical ensemble averages of random networks. Therefore, we present several algorithms to generate random networks conditional on different correlation classes. 

\subsection{The reciprocity of real networks}
All the analytical expressions summarized in Table~\ref{tab:analytical} can be directly estimated by measuring the necessary statistics out of a real network. For different cases specific quantities need to be counted, $N(k_i)$, $N(k_o)$, $N(\boldsymbol{k})$ or $L(k_o \to q_i)$, $L(k_i \to q_o)$, $L(\boldsymbol{k} \to \boldsymbol{q})$, etc. Then, the frequencies like $\mathcal{P}( \boldsymbol{k} \to \boldsymbol{q}) = L(\boldsymbol{k} \to \boldsymbol{q})/L$ or $P(k_i) = N(k_i)/N$ are introduced in the formulas of Table~\ref{tab:analytical} to obtain the expected reciprocity under desired combination of 1-node and 2-node degree correlations.

For each of the real networks in Table~\ref{tab:realnets} we have calculated the expected reciprocities under different correlation classes, Eqs.~(\ref{eq:Genrprob}), (\ref{eq:2noutin}), (\ref{eq:1n}) and (\ref{eq:2noutout}). For comparison, we also show the density of connections $\bar{a}$, i.e., the reciprocity of complete random networks of the same size $N$ and number of links $L$.
As observed in Table~\ref{tab:realnets}, our general estimation of reciprocity $r_{1n2n}$ alone can almost completely explain $r$ of many networks, e.g. World-Trade-Webs and most of the food webs analyzed. It also makes a very good approximation for cortical and neural networks. However, the large discrepancy in the case of Wikipedia web-sites suggests the presence of additional internal structure in the network rather than 1-node and 2-node degree correlations. For example, the \emph{C. elegans} neural network, the cortical and the Wikipedia networks are known to have modular and hierarchical structure. In all these cases the real $r$ is larger than the expected $r_{1n2n}$. Nevertheless, our results demonstrate that degree correlations are crucial to understand the reciprocity in real networks. In its closest approximation, World Trade Webs, $r_{1n2n}$ is roughly twice as large as $\bar{a}$, the expected reciprocity of an equivalent complete random network. In other cases $r_{1n2n}$ and $\bar{a}$ differ up to $3$ orders of magnitude.

\begin{table}
\centering
\begin{tabular}{l c@{\quad}c@{\quad}c@{\quad}c@{\quad}c}
\hline \hline
Network & $r$ & $r_{1n2n}$ & $r_{2n:o/o}$ & $r_{1n}$ & $\bar{a}$ \\ \hline
\multicolumn{6}{l}{World Trade Webs~\cite{wtw}}  \\
Year 1948 & $0.823 $ & $0.812$ & $0.768$ & $0.707$  & $0.382$ \\
Year 2000 & $0.980 $  & $0.958$ & $0.883$ & $0.813$ & $0.560$ \\ \hline
\multicolumn{6}{l}{Neural Networks} \\
C. Elegans~\cite{Elegans} & $0.433$ & $0.329$ & $0.071$ & $0.060$ & $0.033 $ \\ \hline
\multicolumn{6}{l}{Cortical Networks} \\
Cat~\cite{catcortex} & $0.734$ & $0.659$ & $0.473$ & $0.390$ & $0.300$ \\
Macaque~\cite{Macaque} & $0.750$ & $0.645$ & $0.287$ & $0.230$ & $0.155$ \\ \hline
\multicolumn{6}{l}{Food Webs~\cite{foodwebs}} \\
Little Rock lake & $0.0339$ & $0.0323$ & $0.0365$ & $0.0501$ & $0.0743$ \\
Grassland & $0.0 $ & $0.0$ & $0.0077$ & $0.0079$ & $0.0179 $ \\
St. Marks sea. & $0.0$ & $0.0075$ & $0.0500$ & $0.0703$ & $0.0948$ \\
St. Martin Isl. & $0.0$ & $0.0016$ & $0.0419$ & $0.06765$ & $0.1131$ \\
Silwood Park & $0.0$ & $0.0$ & $0.0002$ & $0.0160$ & $0.0155$ \\
Ythan estuary & $0.0034$ & $0.0050$ & $0.0335$ & $0.0531$ & $0.0330$ \\ \hline
\multicolumn{6}{l}{Wikipedia Website~\cite{Zlatic_Wiki_2006}} \\
Spanish  &  $0.3517$ & $0.1466$ & $0.0322$ & $0.0056$ & $0.0004$ \\ 
Portuguese & $0.3563$ & $0.1207$ & $0.0168$ & $0.0084$ & $0.0004$ \\
Chinese  &  $0.3668$ & $0.1556$ & $0.0256$ & $0.0096$ & $0.0010$ \\ 
\hline \hline
\end{tabular}
\caption{Measured reciprocity $r$ of several real networks, and theoretically expected reciprocities due to different correlation structures. (i) 1-node and all 2-node correlations $r_{1n2n}$. (ii) 2-node out-out $r_{2n:o/o}$ (no 1-node). (iii) 1-node correlations $r_{1n}$  (no 2-node). And (iv) the density of connections $\bar{a}$.} 
\label{tab:realnets}
\end{table}

The values of $r_{1n2n}$ are always followed by the contribution of 2-node out-out correlations $r_{2n:o/o}$, Eq.~(\ref{eq:2noutout}). With the exception of Silwood Park and Ythan Estuary food webs, values of $r_{1n}$, Eq.~(\ref{eq:1n}), are all closer to the real $r$ than the density of links $\bar{a}$. These results reveal a hierarchy of the types of degree correlation with respect to their approximation level for explaining reciprocity.

\subsection{Numerical Corroboration}
Finally, we will proof that our analytical expressions are valid expectation values of the network reciprocity. For the real networks in Table~\ref{tab:realnets}, we have generated ensembles of maximally random networks under different conditions of 1-node and 2-node degree correlations. The ensemble average reciprocities have been compared to the analytical results showing excellent agreement between experimental and theoretical expected values, Fig.~\ref{fig:analytical_real}. The generation of random networks with desired degrees and correlations is not trivial, but we have developed three novel rewiring / generation algorithms for that purpose.  
In the absence of analytical results for other graph measures, e.g. clustering coefficient, average pathlength, etc., the following algorithms are also useful to empirically calculate the expected value of any network measure under desired conditions of degree correlations.

\subsubsection{Random networks with desired degree sequences and \\ 2-node degree correlations, $r_{1n2n}$}

Given a real network of size $N$ and $L$ links, we can measure its degree distribution $N(\boldsymbol{k})$ and the 2-node correlation structure $L(\boldsymbol{k} \to \boldsymbol{q})$. With this information in hand, it is possible to generate a maximally random network with such properties. To an initially empty network of size $N$, its nodes are randomly assigned their final degrees $\boldsymbol{k}$ following the distribution $N(\boldsymbol{k})$. Then, links are introduced at random but following carefully considered steps. First, one source node $s$ is chosen at random. We know that $s$ has been assigned to have final degrees $\boldsymbol{k'}$. Because of the 2-node correlations, $s$ can only connect to nodes with particular degrees $\boldsymbol{q'}$ such that $L(\boldsymbol{k'} \to \boldsymbol{q'}) > 0$. A list of possible target nodes is constructed by taking only those nodes assigned to have final degrees $\boldsymbol{q'}$ where $L(\boldsymbol{k'} \to \boldsymbol{q'}) > 0$. From this list one target node $t$ is chosen at random and the link $s \to t$ is include to the initially empty network. Note that a node with input and output degrees $\boldsymbol{k} = (k_i,k_o)$ can only be eligible $k_o$ times as source and $k_i$ time as target, otherwise the distribution $N(\boldsymbol{k})$ will not be conserved. Thus, if the quantities $L(\boldsymbol{k} \to \boldsymbol{q})$, and $k_i$ and $k_o$ of each node are adequately updated during the process, only $L$ iterations are required to construct the maximally random network. Described as it is, this method allows for the introduction of self-loops and multiple-links. Avoiding them is far from trivial.

\begin{figure}  
  \centering
  \includegraphics[width=0.48 \textwidth,clip=]{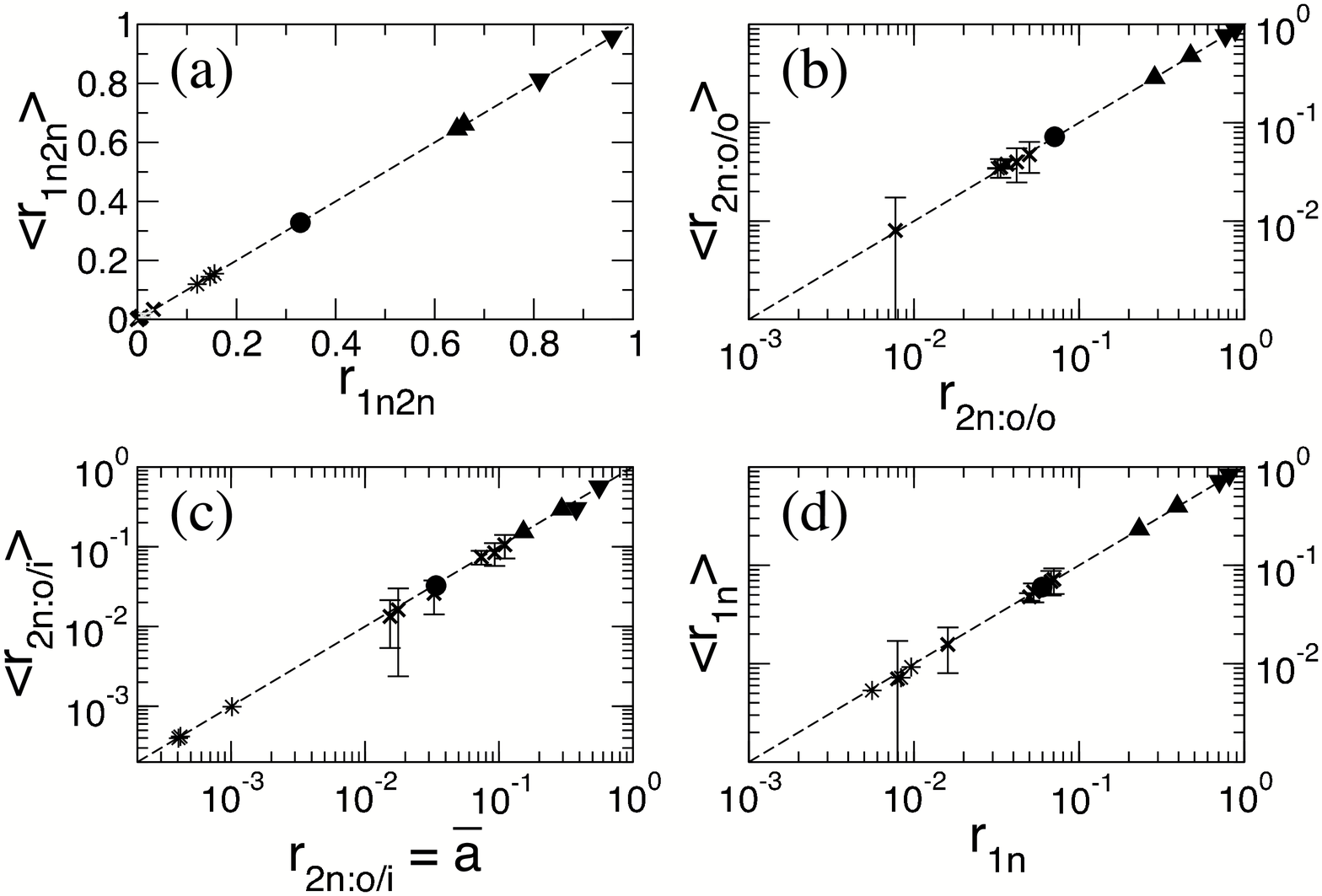}
  \caption {Numerical corroboration of theoretical estimations. Experimentally measured reciprocities (vertical axes) and our theoretical estimations (horizontal axes). a) All 1-node and 2-node correlations conserved. b) Only 2-node out-out correlations conserved.  c) Only 2-node out-in correlations. d) 1-node correlations.  All data points are averages of $100$ realizations.  Food webs ($\times$), C. Elegans ($\bullet$), WTW ($\blacktriangledown$), cortical networks ($\blacktriangle$) and Wikipedias ($\ast$). Dashed lines are references of perfect coincidence.
  \label{fig:analytical_real}} 
\end{figure}

To proof the validity of our general theoretical result $r_{1n2n}$, Eq.~(\ref{eq:Genrprob}), ensembles of $100$ random networks have been generated for each of the real networks in Table~\ref{tab:realnets} following the method described above. The results in Fig.~\ref{fig:analytical_real}(a) show an excellent agreement between $r_{1n2n}$ (horizontal axis) and the empirical ensemble average reciprocities $\left< r_{1n2n} \right>$ (vertical axis).  All the generated random realizations have been positively tested to have the same $N(\boldsymbol{k})$ and $L(\boldsymbol{k} \to \boldsymbol{q})$ as the original real networks. From all the networks in Table~\ref{tab:realnets} none but the C. Elegans contain self-loops and multiple-links. Therefore, the impact of the self-loops and multiple-links introduced by the random generation method was also tested. In the random realizations out of the Silwood Park food web, on average, only $4.5\%$ of the $L$ links formed self-loops or multiple-links. The realizations out of the Wikipedias contain on average less than $2\%$ of such links, and the realizations out of the cortical networks and the world-trade-webs contain less that $1\%$. These small differences have very little impact as observed in the good agreement hown in Fig.~\ref{fig:analytical_real}(a) between our theoretical estimation $r_{1n2n}$ and the experimental measures $\left< r_{1n2n} \right>$.

\subsubsection{Rewiring method that conserves 2-node out-out correlations, $r_{2n:o/o}$}

In this case, we opted for a rewiring algorithm. Such methods start from a given real network and stepwise randomize its connections under specific rules that conserve desired properties. In order to obtain maximally random networks that conserve uniquely the original 2-node out-out degree correlations, Fig.~\ref{fig:knn}(b), we proceed in the following manner. One of the links in the network is chosen at random $s \to t_1$. Nodes $s$ and $t_1$ have out-degrees $k_o$ and $q_o$ respectively. From all the nodes in the network with out-degree $q_o$ a new target node $t_2$ is randomly chosen. If all conditions to avoid the introduction of self-loops and multiple links are satisfied, the old link is destroyed $s \nrightarrow t_1$ and the new link is included $s \to t_2$.

After several iterations, all 2-node degree correlations except for the out-out will be randomized because the rewiring step is blind to all other correlations. A relevant question arises when applying rewiring algorithms: \emph{how long should the process run so that resulting networks are maximally random?} After some finite number of iterations the network reaches a maximally random state and any successive rewiring will lead to an statistically equivalent random network. Once this state is reached, all network measures converge to an stable value. We have applied several levels of rewiring to the real networks studied in this paper and the reciprocity has been measured at each level. As observed in Fig.~\ref{fig:stability}(a) the average values of $r$ reach a stable point. Any other network measure will follow the same behavior. Obviously, the number of necessary iterations is proportional to the number of links $L$ and, for this particular method, all networks reach a maximally random state after $2L$ iterations. 

To proof the validity of our theoretical expression $r_{2n:o/o}$, Eq.~(\ref{eq:2noutout}), as expected reciprocity of networks with prescribed 2-node out-/out-degree correlations, we generated ensembles of $100$ rewired networks out of the real networks in Table~\ref{tab:realnets}. For security, $4L$ rewiring iteration steps were used in all cases.  The ensemble average reciprocities $\left< r_{2n:o/o} \right>$ were calculated. The comparison between our theoretical estimates and the empirical results, Fig.~\ref{fig:analytical_real}(b), shows again excellent agreement.

\subsubsection{Rewiring method that conserves 2-node out-in correlations, $r_{2n:o/i}$}

In order to proof the validity of our theoretical $r_{2n:o/i}$, we designed yet another rewiring algorithm that conserves uniquely the original 2-node out-in correlations Fig.~\ref{fig:knn}(a) of a real network. First, two nodes are selected at random, $s_1$ and $s_2$. These nodes are divided in two halves, one containing all in-links and the other containing all the out-links, i.e. $s_1 = s_1^{in} \, \cup \, s_1^{out}$ and $s_2 = s_2^{in} \, \cup \, s_2^{out}$. Then, the in-halves and the out-halves are switched forming two new nodes, $s_3 = s_1^{in} \cup s_2^{out}$ and $s_4 = s_2^{in} \cup s_1^{out}$. Cases that would introduce self-loops are carefully discarded. The resulting randomized networks conserve the in-degree $N(k_i)$ and the out-degree $N(k_o)$ distributions,  and the number of links $L(k_o \to q_i)$ from the original network while the rest of correlations are randomized. 

\begin{figure}  
  \centering
  \includegraphics[width=0.48 \textwidth,clip=]{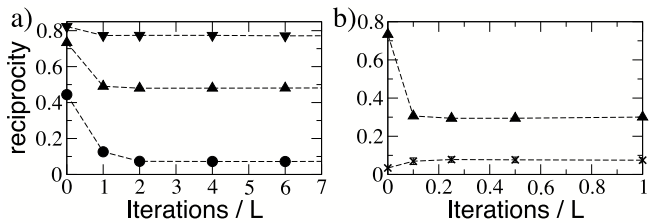}
  \caption {Stability of rewiring algorithms. With increasing number of randomizing iterations structural properties reach stable values, here $r$ is shown. a) Algorithm that conserves the 2-node out-/out-degree correlations. b) Algorithm that conserves the 2-node out-/in-degree correlations. All data points are averages of $50$ realizations. Error bars are very small. Food webs ($\times$), C. Elegans ($\bullet$), WTW ($\blacktriangledown$), cortical networks ($\blacktriangle$).
  \label{fig:stability}} 
\end{figure}

The stability of the rewiring process has been tested in a similar manner as in the previous section and the results are shown in Fig.~\ref{fig:stability}(b). In this case, each iteration step rewires several links so that the algorithm is much faster. After only $0.3L$ iterations all networks reach a maximally random state. The theoretical expected reciprocity $r_{2n:o/i}$, Eq.~(\ref{eq:2noutin}), is compared to the empirical ensemble averages $\left< r_{2n:o/i} \right>$ of $100$ rewired networks and the results shown in Fig.~\ref{fig:analytical_real}(c). All networks were rewired $0.4L$ times.

\subsubsection{Conserving degree sequences, $r_{1n}$}

Finally, a well known rewiring method was used to randomize all 2-node correlations while degree sequences $N(\boldsymbol{k})$, 1-node correlations, are conserved~\cite{Shen-Orr}. This method arises from earlier approaches~\cite{Rao, Roberts} and consists in randomly choosing two links, $s_1 \to t_1$ and $s_2 \to t_2$, and exchanging them, $s_1 \to t_2$ and $s_2 \to t_1$. In Fig.~\ref{fig:analytical_real}(d) the expected reciprocities $\left< r_{1n} \right>$ from the ensemble averages are compared to our analytical estimations $r_{1n}$, Eq.~(\ref{eq:1n}). 

\bigskip
All the results in this section proof that our analytical expressions are valid expected reciprocities under the correlation conditions considered in this paper. The equations are valid even for small networks such as the cat cortical network ($N = 53$, density of links $\bar{a} \approx 0.3$) despite the fact that formulas are derived in the thermodynamical limit. Error bars in most cases are very small and only food webs exhibit some larger fluctuations.

\section{SUMMARY AND DISCUSSION}
In summary, we have studied the influence of 1-node and 2-node degree correlations on the reciprocity of networks with arbitrary degree sequence. We find that, for a large class of complex networks, correlations account for a very large part of the observed reciprocity, explaining it almost completely in typical cases.   In general, the contribution of correlations to $r$ is nontrivial and largely depends on the type of correlations involved, revealing a hierarchy of correlation classes that contribute to $r$ in different levels of approximation. As observed, this influence can span over orders of magnitude in some real networks. Our analytical estimations are proved as valid expectation values of reciprocity by comparison to ensemble averages of random networks which preserve desired types of correlations. Both from a theoretical and a practical point of view, it would be highly desirable to extend the current work and obtain expected values of other network measures, e.g. clustering coefficient, average pathlength, motif profiles, etc. following a similar philosophy: analytical expressions should be computable using only information that can be directly measured from the specific real network under study. In the absence of such results, for the moment, the numerical methods introduced in this paper are a useful tools for significance testing of network measures under conditions of prescribed degree sequences and degree correlations.

The dynamical influence of degree correlations has been studied for epidemic spreading~\cite{Gomez} and for synchronization in undirected networks~\cite{Sorrentino}. In the light of our results, which clearly relate reciprocity and degree correlations, it is important to also investigate the influence of reciprocity on those phenomena for the broader class of directed networks.

In order to perform satisfactory modeling, the key parameters governing network growth and evolution need to be identified. Therefore, theoretical understanding of the interplay between different topological properties is necessary to distinguish between the significant measured values and those expected as by-products of other properties. In this paper, we have explored the interplay between degree correlations and network reciprocity. Previous efforts in this direction include relations between degree correlations and clustering coefficient~\cite{Dorogovtsev_Clustering}. Degree correlations are also expected to be relevant for other network measures, e.g., pathlength, network motifs, modularity, etc. Therefore, further work is desirable to extend the analytical and experimental approaches presented in this paper. No doubt, quantification of similar structural interrelations will significantly elucidate the essence of the structure-function-evolution interplay in complex networks.

\section{ACKNOWLEDGEMENTS}
This work has been supported by the Helmholtz Institute for Supercomputational Physics (G.Z.L.), by GOFORSYS (BMBF) (C.S.Z. and J.K.) and by the Croatian Ministry of Science, Education and Sports under project number 098-0352828-2863 (V.Z. and H.S.)


\end{document}